\documentclass{emulateapj}

\journalinfo{To appear in the Astrophysical Journal \\
{\it submitted 16 September 2009; accepted 25 December 2009}}

\newcommand{\first}{\mbox{\object{HD 209458b}}}
\newcommand{\hdOEN}{\mbox{\object{HD 189733b}}}
\newcommand{\hdOFN}{\mbox{\object{HD 149026b}}}

\newcommand{\gjFTSb}{\mbox{GJ 436b}}
\newcommand{\xooneb}{\mbox{XO-1b}}
\newcommand{\hatpseven}{\mbox{HAT-P-7}}
\newcommand{\tresOne}{\mbox{\object[NAME TrES-1]{TrES-1}}}
\newcommand{\tresTwo}{\mbox{\object[GSC 03549-02811]{TrES-2}}}
\newcommand{\tresThree}{\mbox{\object[NAME TrES-3]{TrES-3}}}
\newcommand{\tresFour}{\mbox{\object[GSC 02620-00648]{TrES-4}}}
\newcommand{\tresTwoTwoMass}{\mbox{\object{2MASS J19071403+4918590}}}

\newcommand{\spi}{{\it Spitzer}}

\shorttitle{Planetary Emission from \tresTwo}
\shortauthors{O'Donovan et al.}

\newcommand\lsim{\mathrel{\rlap{\lower4pt\hbox{\hskip1pt$\sim$}}\raise1pt\hbox{$<$}}}
\newcommand\gsim{\mathrel{\rlap{\lower4pt\hbox{\hskip1pt$\sim$}}\raise1pt\hbox{$>$}}}

\begin{document}

\title{Detection of Planetary Emission from the Exoplanet T\lowercase{r}ES-2 using \emph{Spitzer}/IRAC} 

\author{%
Francis~T.~O'Donovan\altaffilmark{1,2}, %
David~Charbonneau\altaffilmark{3}, %
Joseph~Harrington\altaffilmark{4}, %
N.~Madhusudhan\altaffilmark{5}, %	
Sara~Seager\altaffilmark{6}, %
Drake~Deming\altaffilmark{7}, %
Heather~A.~Knutson\altaffilmark{3} %
}

\altaffiltext{1}{NASA Postdoctoral Program Fellow, Goddard Space Flight Center, 8800 Greenbelt Rd, Greenbelt MD 20771}

\altaffiltext{2}{California Institute of Technology, 1200 E.~California Blvd., Pasadena, CA 91125; ftod@caltech.edu}

\altaffiltext{3}{Harvard-Smithsonian Center for Astrophysics, 60 Garden St., Cambridge, MA 02138}

\altaffiltext{4}{Department of Physics, University of Central Florida, Orlando, FL 32816}

\altaffiltext{5}{Department of Physics, Massachusetts Institute of Technology, 77 Massachusetts Ave., Cambridge, MA 02139}

\altaffiltext{6}{Department of Earth, Atmospheric, and Planetary Sciences, Massachusetts Institute of Technology, 77 Massachusetts Ave., Cambridge, MA 02139}

\altaffiltext{7}{Planetary Systems Laboratory, NASA Goddard Space Flight Center, Mail Code 693, Greenbelt, MD 20771}

\begin{abstract}

We present here the results of our observations of \tresTwo\ using the Infrared Array Camera on \spi.
We monitored this transiting system during two secondary eclipses, when the planetary emission is blocked by the star.
The resulting decrease in flux is 0.127\%$\pm$0.021\%, 0.230\%$\pm$0.024\%, 0.199\%$\pm$0.054\%, and 0.359\%$\pm$0.060\%, at 3.6\,$\mu$m, 4.5\,$\mu$m, 5.8\,$\mu$m, and 8.0\,$\mu$m, respectively. 
We show that three of these flux contrasts are well fit by a black body spectrum with \mbox{$T_{\rm eff}=1500$\,K}, as well as by a more detailed model spectrum of a planetary atmosphere. 
The observed planet-to-star flux ratios in all four IRAC channels can be explained by models with and without a thermal inversion in the atmosphere of \tresTwo, although with different atmospheric chemistry.
Based on the assumption of thermochemical equilibrium, the chemical composition of the inversion model seems more plausible, making it a more favorable scenario. 
\tresTwo\ also falls in the category of highly irradiated planets which have been theoretically predicted to exhibit thermal inversions.
However, more observations at infrared and visible wavelengths would be needed to confirm a thermal inversion in this system.
Furthermore, we find that the times of the secondary eclipses are consistent with previously published times of transit and the expectation from a circular orbit.
This implies that \tresTwo\ most likely has a circular orbit, and thus does not obtain additional thermal energy from tidal dissipation of a non-zero orbital eccentricity, a proposed explanation for the large radius of this planet.
\end{abstract}


\section{Introduction}
\label{sec:intro}

There has been a recent dramatic increase in the number of extrasolar planets within 300\,pc whose structures and atmospheric compositions can be probed using the {\it Spitzer Space Telescope} \citep{Werner_Roellig_Low:apjs:2004a}. 
These are the nearby transiting exoplanets. 
By measuring with {\it Spitzer} the decrease in light as one of these planets passes behind its star in an event known as a secondary eclipse, we can estimate the flux emitted by the planet.  
Detection of this emission from planetary atmospheres is made possible by taking advantage of the enhanced contrast between stars and their planets in the infrared wavelengths observable with {\it Spitzer}.
Combining several of these flux measurements allows us to characterize the shape of the planet's emission spectrum, which tells us about the properties of its dayside atmosphere. (\citealp[For a discussion of extrasolar planetary atmospheres, see][]{Charbonneau_Brown_Burrows:PPV:2007a, Marley_Fortney_Seager:PPV:2007a}).

The planets \first\ and \hdOEN\ have been the optimal choices for \spi\ studies of extrasolar planetary atmospheres, because of their early discovery and the relative brightness of their stellar hosts.
We have detected infrared emission from these exoplanets, both photometrically \citep{Deming_Seager_Richardson:nat:2005a, Deming_Harrington_Seager:apj:2006a, Knutson_Charbonneau_Allen:nat:2007a, Knutson_Charbonneau_Allen:apj:2008a}, and with low-resolution spectroscopy  \citep{Richardson_Deming_Horning:Nature:2007a, Grillmair_Charbonneau_Burrows:apjl:2007a, Swain_Bouwman_Akeson:apj:2008a, Charbonneau_Knutson_Barman:apj:2008a, Grillmair_Burrows_Charbonneau:nat:2008a, Swain_Vasisht_Tinetti:Nature:2008a, Swain_Tinetti_Vasisht:apj:2009a, Swain_Vasisht_Tinetti:apjl:2009a}.
Another early success was the report by \cite{Charbonneau_Allen_Megeath:apj:2005a} of the measurement of infrared light from \tresOne. 
More recently, \cite{Harrington_Luszcz_Seager:nat:2007a}, \cite{Deming_Harrington_Laughlin:apjl:2007a}, \cite{Demory_Gillon_Barman:aa:2007a}, and \cite{Machalek_McCullough_Burke:apj:2008a} announced the results from \spi\ observations of the transiting exoplanets \hdOFN, \gjFTSb, and \xooneb.

There has been a flurry of activity to reconcile atmospheric models with this limited number of infrared measurements. 
While several attempts have been made to explain the infrared observations (\citealp[see, e.g.,][]{Barman_Hauschildt_Allard:apj:2005a, Burrows_Hubeny_Sudarsky:apjl:2005a, Burrows_Hubeny_Budaj:apj:2007a, Fortney_Marley_Lodders:apjl:2005a, Fortney_Cooper_Showman:apj:2006a, Seager_Richardson_Hansen:apj:2005a}), the models are not entirely in agreement and no single model can explain every observation. Recently, \cite{Burrows_Hubeny_Budaj:apjl:2007a}, \cite{Burrows_Budaj_Hubeny:apj:2008a}, and \cite{Fortney_Lodders_Marley:apj:2008a} supplied a possible piece of the puzzle by proposing that the very highly irradiated hot Jupiters such as \first\ and \tresTwo\ (see Fig.~1 of \citealt{Fortney_Lodders_Marley:apj:2008a}) will exhibit water emission rather than the expected water absorption at the IRAC wavelengths, a result of a temperature inversion in their atmospheres. The large stellar insolation that \tresTwo\ experiences may permit the presence of TiO and/or VO molecules in the hot planetary atmosphere that would condense in a cooler atmosphere. These opaque molecules would then cause the inversion. 
However, while this hypothesis can explain the emission features seen in the spectrum of \first\ \citep{Burrows_Hubeny_Budaj:apjl:2007a, Knutson_Charbonneau_Allen:apj:2008a}, it is incomplete. 
The relatively lower insolation experienced by the planet \xooneb\ is similar to that of \tresOne, and hence both atmospheres would be predicted to have water absorption features, as is indeed supported by the infrared observations of \tresOne\ by \cite{Charbonneau_Allen_Megeath:apj:2005a}. 
Nevertheless, \cite{Machalek_McCullough_Burke:apj:2008a} have shown that \xooneb\ displays the contrary, with evidence of water emission in its atmosphere. More recently, \cite{Fressin_Knutson_Charbonneau:preprint:2009a} did not find any evidence for the expected thermal inversion in the atmosphere of the highly irradiated planet \tresThree.
\cite{Zahnle_Marley_Freedman:apjl:2009a} have shown that sulfur photochemistry may also lead to the formation of inversions. 
In their models, this photochemistry is more or less temperature independent between 1200\,K and 2000\,K, and is also relatively insensitive to atmospheric metalicity.

Although over 400 extrasolar planets are known, it is only for these nearby transiting exoplanets that we can measure the planetary radii and true planetary masses precisely enough to provide useful constraints for theoretical models. 
There have been problems reconciling the observed planetary masses and radii with models (see \citealp{Laughlin_Wolf_Vanmunster:apj:2005a}, \citealp{Charbonneau_Brown_Burrows:PPV:2007a}, \citealp{Liu_Burrows_Ibgui:apj:2008a}, and \citealp{Ibgui_Burrows:apj:2009a}, for a review), namely that there are planets such as \tresTwo\ whose radii are larger than predicted by the standard models for an irradiated gas giant planet. 
Several explanations have been proposed for bloated planets like \tresTwo, mainly related to some additional source of energy to combat planetary contraction.
One such possible energy source is the tidal damping of a non-zero eccentricity \citep{Bodenheimer_Lin_Mardling:apj:2001a, Bodenheimer_Laughlin_Lin:apj:2003a}: the planetary orbit of a hot Jupiter is expected to be circular, unless it is gravitationally affected by an unseen planetary companion (\citealp[see, e.g.,][]{Rasio_Ford:science:1996a}). 
We can constrain the likelihood of a tidal damping energy source by measuring the timing of the secondary eclipse of a transiting hot Jupiter like \tresTwo\ and comparing these timings to predictions based on transit timings and the hypothesis of a circular planetary orbit (\citealp[see, e.g.,][]{Deming_Seager_Richardson:nat:2005a}). 

The transiting hot Jupiter \tresTwo\ \citep{ODonovan_Charbonneau_Mandushev:apjl:2006a} is one of the known hot Jupiters with a radius larger than expected from current models. 
The atmosphere of \tresTwo\ experiences similar levels of irradiation from its host star as the atmosphere of the likewise bloated planet \first, and hence we expect to find evidence of a thermal inversion for \tresTwo, according to the predictions of \cite{Burrows_Hubeny_Budaj:apjl:2007a}, \cite{Burrows_Budaj_Hubeny:apj:2008a}, and \cite{Fortney_Lodders_Marley:apj:2008a}. 
Here we present the first \spi\ observations of \tresTwo\ (\S\ref{sec:obs}). 
From our analysis (\S\ref{sec:lc}), we have detected thermal emission from the transiting planet. We found no evidence for timing offsets of the secondary eclipses, and deduce the possible presence of a thermal inversion in the atmosphere of \tresTwo\ (\S\ref{sec:discuss}). 

\section{IRAC Observations of TrES-2}
\label{sec:obs}

We monitored \tresTwo\ using \spi\ during the time of two secondary eclipses, employing a different pair of the four bandpasses available on the Infrared Array Camera (IRAC; \citealt{Fazio_Hora_Allen:apjs:2004a}) during each of the eclipses. 
We took care to position \tresTwo\ (\tresTwoTwoMass: \mbox{$J=10.232$\,mag}, \mbox{$J-K_{s}=0.386$\,mag}) away from array regions impaired by bad pixels or scattered light. 
We also kept the corresponding IRAC stray light avoidance zones free of stars that are bright in the infrared. 
We observed a \mbox{$5\farcm2 \times 5\farcm2$} field of view (FOV) containing \tresTwo\ during two eclipses, using the Stellar Mode of the IRAC instrument.
On UT 2006 November 30 (starting at \mbox{$\mathrm{HJD}\,2,\!454,\!069.956$}), we obtained 1073 images of this FOV at 4.5\,$\mu$m and 8.0\,$\mu$m with an effective integration time of 10.4\,s for a total observing time of 3.9\,hr.
Our 3.6-$\mu$m and 5.8-$\mu$m observations of \tresTwo\ were taken on UT 2007 August 16.
The observations began at \mbox{$\mathrm{HJD}\,2,\!454,\!324.436$} and lasted 4.0\,hr, during which we acquired 2130 and 1065 images in the respective channels (of effective exposure time 1.2s and 10.4s, respectively).

\section{Deriving and Modeling Light Curves of TrES-2}
\label{sec:lc}

\begin{deluxetable}{lcc}
\tablewidth{0pt}
\tablecaption{TrES-2 System Parameters \label{tab:tres2}}
\tablehead{ \colhead{Parameter} & \colhead{Value}  &  \colhead{Reference} }
\startdata

$P$ \phm{000.} (d) &  \phm{000000} $2.470621\pm0.000017$ & a \\
$T_{c}$ \phm{.} (HJD)  & $2,\!453,\!957.63479\pm0.00038$ & a \\
$b=a \cos{i} / R_{\star}$ & \phm{000000.}$0.8540 \pm 0.0062$ & a \\ 
$i$ \phm{00000} ($\degr$)  &  \phm{000000} $83.57\pm0.14$  & a \\
$R_{p}/R_{\star}$ &  \phm{000000} $0.1253 \pm 0.0010$ & a \\ 

& & \\

$T_{s}$ [3.6\,$\mu$m/5.8\,$\mu$m]  (HJD)  & $2,\!454,\!324.5220\pm0.0026$ & b \\
$T_{s}$ [4.5\,$\mu$m/8.0\,$\mu$m]  (HJD)  & $2,\!454,\!070.04805\pm0.00086$ & b \\

\enddata
\tablenotetext{a}{\cite{Holman_Winn_Latham:apj:2007a}.}
\tablenotetext{b}{This work.}
\end{deluxetable}

As part of the Spitzer Science Center pipeline for IRAC data (version S15.0.5 for the 2006 observations, version S16.1.0 for the 2007 observations), the images were corrected for dark current, flat-field variations, and certain detector nonlinearities. 
Each header of these Basic Calibrated Data (BCD) images contains the time and date of observation and the effective integration time. 
We used these to compute the Julian date corresponding to mid-exposure of each observation. 
In order to convert these dates to Heliocentric Julian dates, we calculated the corresponding light travel time between the Sun and \spi\ (ranging from 0.5 to 2.5\,minutes), using NASA JPL's HORIZONS%
\footnote{\url{http://ssd.jpl.nasa.gov/?horizons}}%
\ service.
We then computed the orbital phases using an updated orbital period and transit epoch (see Table~\ref{tab:tres2}) derived from observations \citep{Holman_Winn_Latham:apj:2007a} made as part of the Transit Light Curve project. 

\begin{figure*} % Make one-column
\epsscale{0.9}
\plotone{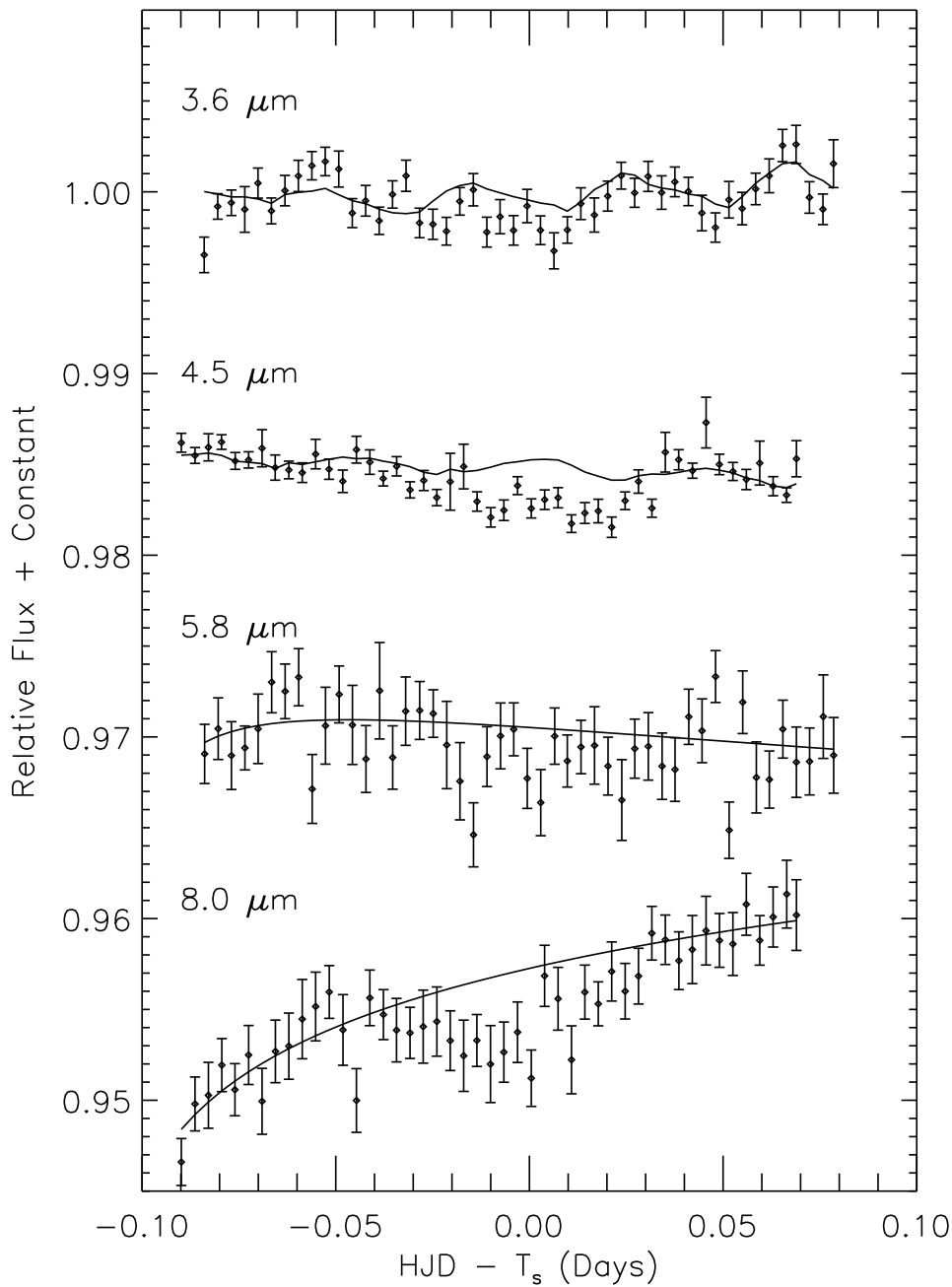}
\caption{%
Relative fluxes from the system TrES-2 at 3.6\,$\mu$m, 4.5\,$\mu$m, 5.8\,$\mu$m, and 8.0\,$\mu$m (with an arbitrary flux offset), binned and plotted vs.\ the time from the predicted center of secondary eclipse ($T_{s}$). 
Superimposed are our best-fit models (\emph{black lines}) for known instrumental effects.
The error bar shown for each binned data point is the standard deviation of the flux values in that bin, divided by the square root of the number of points in the bin. 
}
\label{fig:tres2before}
\end{figure*}

Using the nearest integer pixel $(x_{0}, y_{0})$ as an initial estimate of the position of \tresTwo\ on the array, we computed the flux-weighted centroid $(x, y)$ of \tresTwo\ in each BCD image. 
The intra-pixel position of \tresTwo\ is then $(x' = x - x_{0}, y' = y - y_{0})$.
We measured the flux from our target using circular apertures ranging from 2 to 10 pixels, and subtracted the background signal assessed in a sky annulus with inner and outer radii of 20 and 30 pixels, respectively. 
We normalized the fluxes for a given channel and aperture size by dividing each time series by the median of its values outside the times of eclipse. 
We examined the variation of the residual scatter of the out-of-eclipse data, and found that an aperture radius of 3 pixels produced the smallest residual scatter for all four channels. 
Figure~\ref{fig:tres2before} shows the four light curves we obtained using this photometric aperture. 

For each of the observed secondary eclipses, we created a model of the two IRAC light curves obtained during that eclipse. 
We first accounted for various detector effects known to be present in IRAC data. 
There is a known correlation between the IRAC 3.6-$\mu$m or 4.5-$\mu$m flux from a source and the intra-pixel position $(x', y')$ on the detector \citep{Reach_Megeath_Cohen:pasp:2005a, Charbonneau_Allen_Megeath:apj:2005a, Charbonneau_Knutson_Barman:apj:2008a, Knutson_Charbonneau_Allen:apj:2008a}: the sensitivity of an individual pixel varies depending on the location of the stellar point-spread function (PSF), with higher fluxes measured near the center of the pixel and lower fluxes near the edges.
Figure~\ref{fig:tres2before} shows this effect in our \tresTwo\ data. 
Data from these two channels also demonstrate a linear trend with time 
(as previously observed by \citealp{Knutson_Charbonneau_Burrows:apj:2009a} in IRAC observations of \tresFour), with a positive trend at 3.6\,$\mu$m and a negative slope at 4.5\,$\mu$m.
We removed these trends to obtain $f_{0}$, the actual stellar flux:
\begin{equation}
f_{0} =  f / ([c_{1} + c_{2}x' + c_{3}y' + c_{4}(x')^2 + c_{5}(y')^2] \times [1 + Cdt]),
\end{equation}
where $f$ is the measured stellar flux, $(x', y')$ is the intral-pixel position, $dt$ is the amount of time from the first observation, and $(c_{1\mbox{-}5}, C)$ are free parameters in our model.
The 5.8-$\mu$m and 8.0-$\mu$m data (see Fig.~\ref{fig:tres2before}) display the ``ramp'' associated with these IRAC channels first noticed by \cite{Charbonneau_Allen_Megeath:apj:2005a}, and expanded upon by \cite{Harrington_Luszcz_Seager:nat:2007a} (supplementary information). 
Both data sets showed an overall increase in flux with $dt$.
For these two data sets, we removed this detector effect by including the following correction in the model:
\begin{equation}
f_{0} = f / [d_{1} + d_{2}(\ln{dt'}) + d_{3}(\ln{dt'})^2],
\end{equation}
where $d_{1\mbox{-}3}$ are free parameters, and $dt'=dt + 0.02$\,days.
Here our substitution of $dt'$ for $dt$ is simply to avoid division by infinity. 
We then modeled simultaneously the two corrected time series using the eclipse light curve equation for a uniform source from \cite{Mandel_Agol:apjl:2002a}.
We obtained the required system parameters (see Table~\ref{tab:tres2}) from \cite{Holman_Winn_Latham:apj:2007a}: the planetary orbital period, impact parameter, orbital inclination, and the radius ratio between the planet and the star. 
Based on this ephemeris, we calculated the predicted eclipse epoch ($T_{s}$; see Table~\ref{tab:tres2}). 
The three free parameters for the eclipse model were the timing offset ($\Delta t$), the eclipse depth at the shorter wavelength ($\Delta f_{l}$), and the depth at the longer wavelength ($\Delta f_{h}$). 
The model of the two light curves therefore has 12 free parameters: [$c_{1}$, $c_{2}$, $c_{3}$, $c_{4}$, $c_{5}$, $C$, $d_{1}$, $d_{2}$, $d_{3}$, $\Delta t$, $\Delta f_{l}$, $\Delta f_{h}$].

For a particular instance of this model, we compute the $\chi^2$ measure of the goodness of fit to the two relevant data sets as follows.
For each light curve, we compute an initial uncertainty for the normalized flux values as their standard deviation. 
We exclude 5-$\sigma$ outliers in flux from further consideration.
We then exclude large outliers in intra-pixel position ($\max{[|x' - x'_{m}|, |y' - y'_{m}|]} > 0.15$, where [$x'_{m}, y'_{m}$] are the median intra-pixel values) on the detector.
We compute the $\chi^2$ of the two light curves separately, and rescale each $\chi^2$ to reflect the reduction of the number of data points by the above outlier exclusion. 
We then sum the resulting values to compute the $\chi^2$ of the overall model. 

To find an initial estimate of the best-fit parameters for each model of two light curves, we first used the AMOEBA algorithm \citep{Press_Teukolsky_Vetterling:1992a} to minimize the $\chi^2$ of the fit. 
Using this initial estimate as a starting point, we applied the Markov Chain Monte Carlo method \citep[see, e.g.,][]{Ford:aj:2005a, Winn_Holman_Fuentes:aj:2007a}, computing the $\chi^2$ at each of the $10^6$ steps of the chain.
We then calculated the median of the $10^6$ $\chi^2$ values, and excluded from further analysis all the steps prior to the occurrence of the first value lower than this median.
For the $i$th free parameter of our model, we derived the best-fit value $p_{i}$ as the median of the remaining distribution of values for that parameter. 
We computed the (possibly unequal) lower ($\sigma_{-,i}$) and upper ($\sigma_{+,i}$) errors in this value such that the ranges $[p_{i}-\sigma_{-,i},p_{i}]$ and $[p_{i},p_{i}+\sigma_{+,i}]$ each contain 68\%/2 of the values less than or greater than, respectively, the best-fit value. 
We computed the flux residuals after dividing out the best-fit model, and then derived an updated uncertainty as the new standard deviation of the residuals from the model.
We again computed and rescaled the $\chi^2$ of the two light curves separately, and then summed the resulting values to compute the $\chi^2$ of the best-fit. 

\begin{deluxetable}{lc}
\tablewidth{0pt}
\tablecaption{Best-Fit Values for Free Parameters of Eclipse Models\label{tab:bestfit}}
\tablehead{ \colhead{Parameter} & \colhead{Value} }
\startdata

& \\
3.6\,$\mu$m/5.8\,$\mu$m & \\
& \\

$\chi^{2}$ & 3164\\
$N$ & 3166\\
$\Delta t$ (min) & $1.8\pm3.6$\\
$\Delta f_{l} (= \Delta f_{3.6\,\mu m}$) & 0.127\%$\pm$0.021\%\\
$\Delta f_{h} (= \Delta f_{5.8\,\mu m}$) & 0.199\%$\pm$0.054\%\\

& \\
4.5\,$\mu$m/8.0\,$\mu$m & \\
& \\

$\chi^{2}$ & 2117 \\
$N$ & 2119\\
$\Delta t$ (min) & $0.7\pm$3.1\\
$\Delta f_{l} (= \Delta f_{4.5\,\mu m}$) & 0.230\%$\pm$0.024\%\\
$\Delta f_{h} (= \Delta f_{8.0\,\mu m}$) & 0.359\%$\pm$0.060\%\\

\enddata
\end{deluxetable}

 % Make one-column

\begin{figure*} % Make one-column
\epsscale{0.9}
\plotone{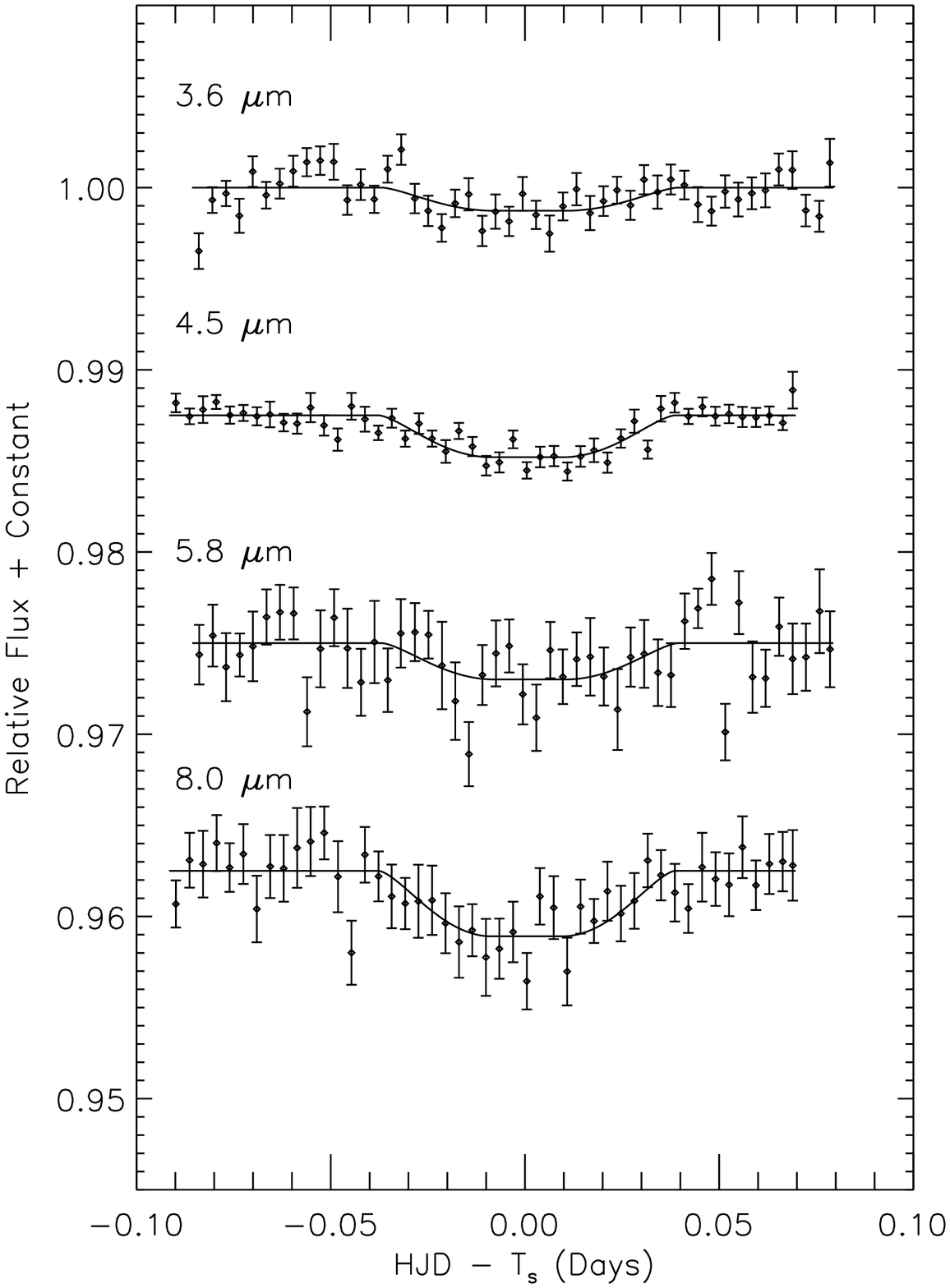}
\caption{%
Same relative fluxes as Figure~\ref{fig:tres2before}, except that here the fluxes have been corrected for the known detector effects. 
Overplotted are our best-fit models (\emph{black lines}) of the secondary eclipses.
The error bars are as in Figure~\ref{fig:tres2before}. 
}
\label{fig:tres2after}
\end{figure*} % Make one-column

We tabulate for both pairs of channels the best-fit values for the free parameters of the eclipse models and the reduced $\chi^{2}$ in Table~\ref{tab:bestfit}.
We have overplotted in Figure~\ref{fig:tres2before} the above corrective functions using the best-fit parameters we derived from the Markov chains. 
In Figure~\ref{fig:tres2after}, we plot the corrected fluxes from \tresTwo\ at the four wavelengths, and overplot the best-fit eclipse models. 
In both figures, the error bar shown for each binned data point is the standard deviation of the flux values in that bin, divided by the square root of the number of points in the bin. 

\section{Discussion and Conclusions}
\label{sec:discuss}

From the best-fit values (see Table~\ref{tab:bestfit}) for the timing offset of the two observed secondary eclipses, we see that their weighted average ($\Delta t = 1.2\pm2.3$\,min) is consistent with no offset from the predicted epochs for the eclipses. 
Note that we have not accounted for the light travel delay time \citep[see][]{Loeb:apjl:2005a} of 37\,s across the \tresTwo\ system, because of the relatively large size of the errors in these timing offsets compared to this delay time. 
An upper limit for the orbital eccentricity of a transiting planet can be computed from the timing offset $\Delta t$, using $e \, \cos{\omega} \simeq \pi \, \Delta t/2 \, P$, where $\omega$ is the unknown longitude of periastron and $P$ is the known orbital period (\citealp[][ Eq.~4]{Charbonneau_Allen_Megeath:apj:2005a}). 
The 3-$\sigma$ upper limit for \tresTwo\ is therefore $0.0036$, consistent with a negligible orbital eccentricity, unless $w\simeq90\degr$.
Tidal damping of orbital eccentricity \citep{Bodenheimer_Lin_Mardling:apj:2001a, Bodenheimer_Laughlin_Lin:apj:2003a} is therefore unlikely to be a sizable contribution to the internal energy of this bloated exoplanet. 

\begin{figure*} % Make one-column
\includegraphics[scale=1.00]{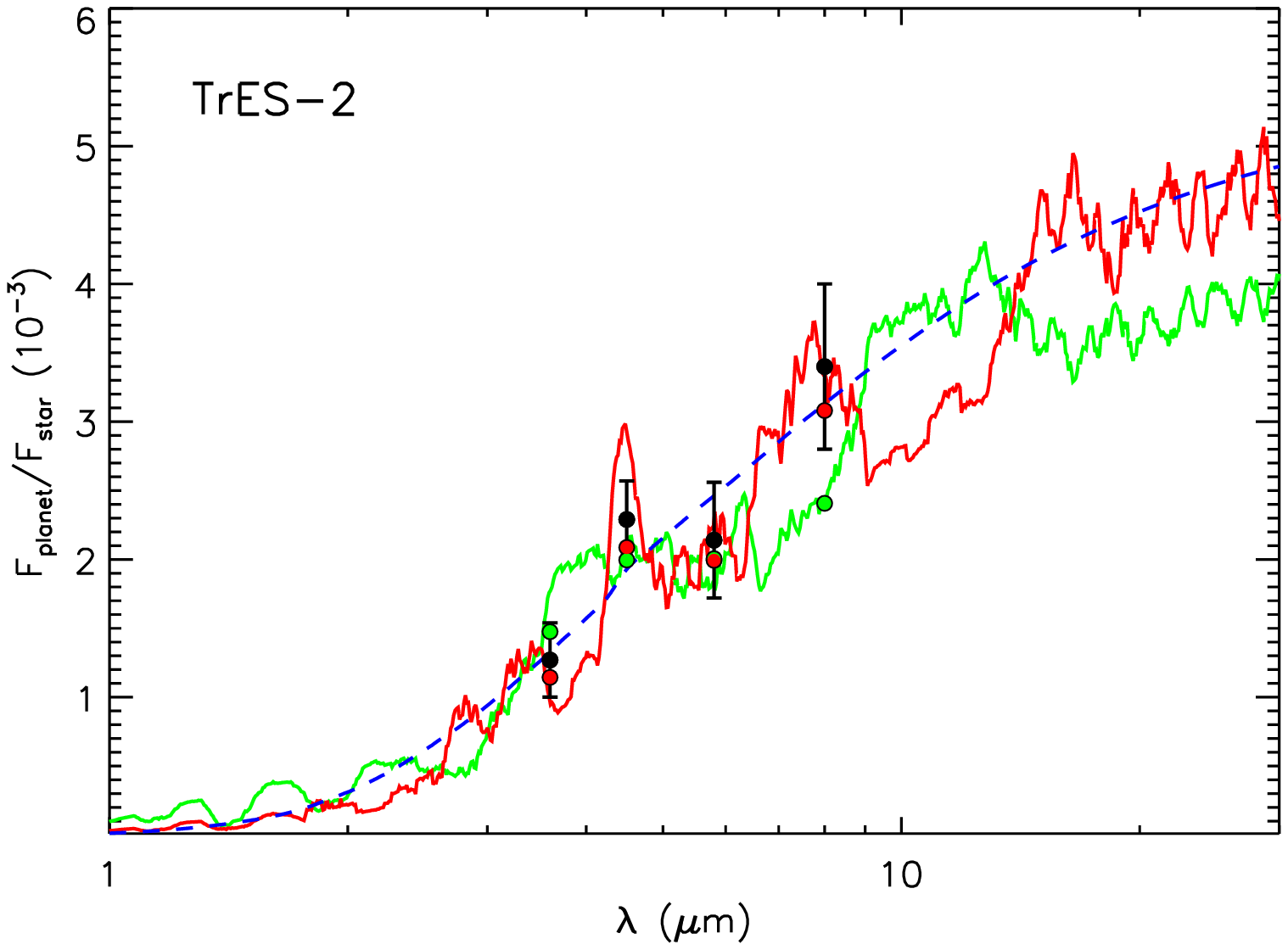}
\caption{%
Contrast ratios ({\it black filled circles}) for TrES-2 at 3.6\,$\mu$m, 5.8\,$\mu$m and 8.0\,$\mu$m, which are consistent with a model ({\it blue line}) of a 1500\,K black body planetary flux divided by a Kurucz model of the star TrES-2. Our additional observation at 4.5\,$\mu$m shows some evidence of excess emission at this wavelength. Also shown are the predictions ({\it red and green circles}) for these four fluxes from theoretical planet-star flux contrast models ({\it red and green lines}) computed for the star TrES-2 using the \cite{Madhusudhan_Seager:apj:2009a} code (see text). The {\it red} ({\it green}) spectrum corresponds to a model with (without) a thermal inversion on the planet dayside.%
}
\label{fig:spectra}
\end{figure*} % Make one-column

We now turn to a discussion of the \tresTwo\ planet-star contrasts. We first emphasize how well the data can be fit by a blackbody spectrum with no molecular band features. The black filled circles with error bars in Figure~\ref{fig:spectra} show the {\it Spitzer} IRAC data points from this work. The {\it blue} dashed line is a 1500\,K blackbody flux divided by the Kurucz stellar model  with stellar parameters ($T_{\rm eff}=5750$\,K, $Z=0.0$, $\log(g)=4.5$) closest to those ($T_{\rm eff}=5850\pm50$\,K, $Z=-0.15\pm0.10$, $\log(g)=4.4\pm0.1$) derived by \cite{Torres_Winn_Holman:apj:2008a}. It is clear that the blackbody fits all the data points resonably well, given the large error bars. However, a blackbody spectrum is only a nominal guideline, since the actual planetary spectrum is influenced by the myriad contributions due to molecular band features, collision induced opacities, temperature gradients.

\begin{figure*} % Make one-column
\includegraphics[scale=1.00]{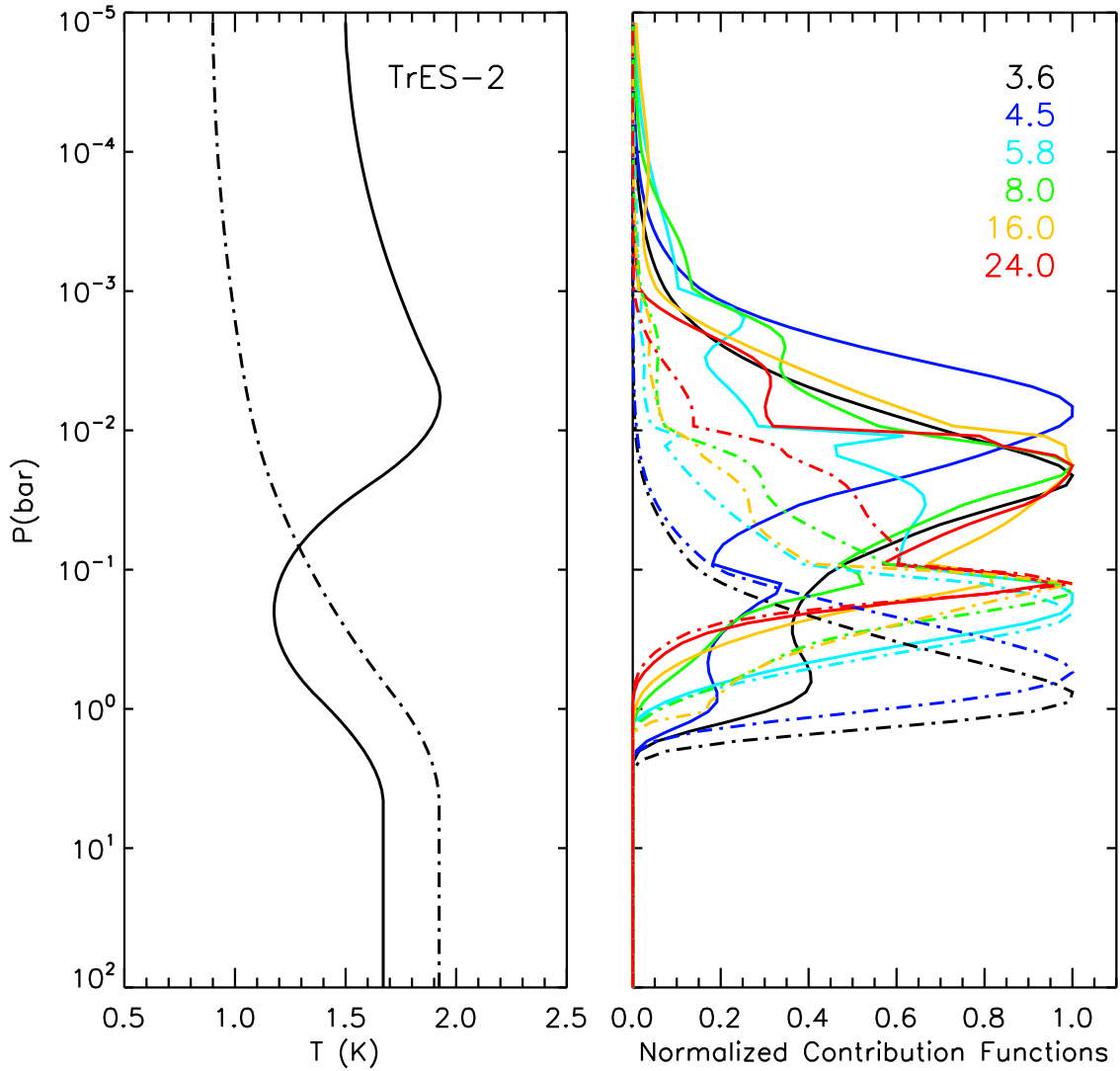}
\caption{%
Pressure-Temperature profiles and contribution
functions corresponding to the models shown in Figure~\ref{fig:spectra}.
The left panel
shows the two pressure temperature profiles (with and without
a thermal inversion) corresponding to the two models
reported in this work. The {\it solid} ({\it dash-dot}) curve shows
the P-T profile with (without) a thermal inversion. Both the profiles lead to
good fits to the observations, albeit with different molecular
compositions. The $P$-$T$ profiles were generated with the parametric
prescription developed in \cite{Madhusudhan_Seager:apj:2009a}. 
The thermal
inversion in the {\it solid} profile lies between 0.20--0.01\,bar,
and spans temperatures between 1100--2000\,K.
The right panel shows the contribution functions in the six
Spitzer photometric channels for each model. The legend
shows the channel center wavelength in microns, and the
curves are color-coded by the channel. The {\it solid} curves
show the contribution functions for the inversion model,
whereas the {\it dash-dot} curves show the contribution
functions for the non-inversion model. All the contribution
functions are normalized to unity.
}
\label{fig:ptprofiles}
\end{figure*}

Model atmospheric spectra for \tresTwo\ are also shown in Figure~\ref{fig:spectra}. The red and green spectra show models with and without a thermal inversion on the planet dayside, respectively. The red and green circles (enclosed in black circles) show the corresponding model points obtained by integrating the spectra over the {\it Spitzer} IRAC bandpasses. The corresponding model thermal profiles are shown in Figure~\ref{fig:ptprofiles}. The spectra were generated using the hot Jupiter atmosphere model developed in \cite{Madhusudhan_Seager:apj:2009a}. We consider a cloudless atmosphere, and the molecular species are assumed to be well mixed. The stellar spectrum was represented by the appropriate Kurucz model. The model spectrum without a thermal inversion has $T_{\rm eff}=1634$\,K, and the model spectrum with a thermal inversion has $T_{\rm eff}=1459$\,K. 
Both the models allow for extremely efficient day-night energy redistribution. 
At face value, we find that both the models fit the data almost equally well. 

However, the two models require different molecular compositions, which helps us determine the more probable model. The model without a thermal inversion has uniform molecular mixing ratios of $10^{-4}$ for H$_2$O, $10^{-6}$ for CO, and $10^{-6}$ for CH$_4$. While the mixing ratio of H$_2$O is plausible, the low mixing ratio of CO is surprising. In a hot atmosphere, with $T_{\rm eff} = 1500$\,K, CO is expected to be highly abundant. On the other hand, the model with a thermal inversion has uniform molecular mixing ratios of $10^{-4}$ for H$_2$O, $10^{-4}$ for CO, $ 5 \times 10^{-5}$ for CH$_4$, and $2 \times 10^{-6}$ for CO$_2$. These compositions show a relatively high abundance of CO, as expected in a hot Jupiter atmosphere like that of \tresTwo. Based on these compositions, the presence of a thermal inversion seems like a more favorable scenario. 
Additionally, the requirement of $2 \times 10^{-6}$ of CO$_2$ for the inversion model suggests an enhanced metallicity of $[M/H]\sim0.7$ \citep{Zahnle_Marley_Freedman:apjl:2009a}.
However, a thorough exploration of the parameter space is needed to place constraints on the thermal inversion in conjunction with the molecular compositions \citep{Madhusudhan_Seager:ApJ-submitted:2010a}. 

Thus we see some evidence that \tresTwo\ is part of the group of planets that display excess emission at wavelengths longer than 4\,$\mu$m.
The presence of a thermal inversion in the atmosphere of \tresTwo\ would support the theory of \cite{Fortney_Lodders_Marley:apj:2008a} and \cite{Burrows_Budaj_Hubeny:apj:2008a} that planets with substellar fluxes greater than or equal to approximately $10^9\,\mathrm{erg\ s^{-1}\,cm^{-2}}$ should show evidence for a temperature inversion in their atmospheres.

Our theoretical models for the planetary flux from \tresTwo\ are currently constrained only at infrared wavelengths.
Such an infrared signal appears to be relatively insensitive to cloud cover \citep{Burrows_Hubeny_Sudarsky:apjl:2005a}.
However, the amount of starlight reflected off the planetary atmosphere (at optical wavelengths) is highly dependent on the presence and size of upper-atmospheric condensates such as MgSiO$_{3}$, Fe and Al$_{2}$O$_{3}$ \citep{Marley_Gelino_Stephens:apj:1999a,Green_Matthews_Seager:apj:2003a}.
Hence, a tighter constraint of atmospheric models for \tresTwo\ would be derived from the combination of measurements of the optical planetary flux from observations of secondary eclipses with the results from our infrared observations.
Two likely sources for such observations of the optical flux from \tresTwo\ are the Kepler mission \citep{Borucki_Koch_Dunham:ASP:1997a} and the EPOXI mission \citep{Deming_AHearn_Charbonneau::2007a}. Kepler has already demonstrated the ability to monitor the variation in thermal emission and reflected light from \hatpseven\ \citep{Borucki_Koch_Jenkins:science:2009a}.
We eagerly await the comparison of our infrared observations with future optical observations from these two missions.

Although our observations of the atmospheric emission from \tresTwo\ at 3.6\,$\mu$m, 4.5\,$\mu$m, 5.8\,$\mu$m, and 8.0\,$\mu$m provide limited spectral coverage, we have been able to deduce the probable presence of gaseous molecules with high opacities in the atmospheric that result in emission in the 4.5\,$\mu$m band.
The highly irradiated gas giant \tresTwo\ thus may provide additional evidence for the correlation between the occurrence of thermal inversion in the atmosphere of a planet and the level of stellar insolation experienced by the planet.

Our observations of two secondary eclipses by this exoplanet occurred at the time predicted using the time of transit and the assumption of a circular orbit, within the errors. From this we conclude that the source of additional energy required to inflate the planetary radius to its bloated size is unlikely to be tidal heating caused by the circularization of an eccentric orbit. 

\acknowledgments	

This work is based on observations made with the Spitzer Space Telescope, which is operated by the Jet Propulsion Laboratory, California Institute of Technology under a contract with the National Aeronautics and Space Administration (NASA). 
This research was supported in part by NASA under grant NNG05GJ29G % Sleuth
(issued through the Origins of Solar Systems Program), and also by an appointment to the NASA Postdoctoral Program at the Goddard Space Flight Center (administered by Oak Ridge Associated Universities through a contract with NASA).

{\it Facilities:} \facility{Spitzer (IRAC)}

 % Make one-column

\bibliographystyle{apj}
\bibliography{apjmnemonic,mybib.planets}

\end{document}